\documentstyle[12pt]{article}
\begin{document}
\baselineskip=24pt
\begin{titlepage}
\begin{center}
{\Large
Models for Monolayers Adsorbed on a Square Substrate
}
\end{center}
\medskip
\medskip
\bigskip
\begin{center}
{\large Ofer Biham, Lee-Wen Chen,
and Gianfranco Vidali}\\
\end{center}
\bigskip
\medskip
\begin{center}
{\large
Department of Physics \\
Syracuse University \\
Syracuse, NY 13244 \\
}
\end{center}
\end{titlepage}

\newpage
\begin{center}
{\bf Abstract}
\end{center}
Motivated by recent experimental studies of Hg and Pb monolayers on
Cu(001) we introduce a
zero temperature model of a monolayer adsorbed on a square substrate.
Lennard-Jones potentials are used to describe the interaction between
pairs of adlayer-adlayer and adlayer-substrate atoms.
We study a special case in which the
monolayer atoms form a perfect square structure and the lattice
constant, position and
orientation with respect to the substrate can vary to minimize the energy.
We introduce a rule based on the Farey tree construction to generate
systematically the most energetically favored phases and use
it to calculate the phase diagram in this model.

\newpage
\section{Introduction}

The study of ordered phases of films of one layer or less on metal substrates
is an active topic of research in two-dimensional physics \cite{shrimpton}.
Due to improved film deposition and
detection techniques, it is now possible to
investigate more systems and with better resolution than has been possible
before \cite{unertl}.
Experimental studies of ordered overlayers show a great variety of commensurate
and incommensurate phases
\cite{bak,roelofs}.
Here we define a commensurate phase as an
ordered structure for  which it is possible to find a primitive cell with basis
vectors which are  linear combinations of lattice vectors of the substrate. If
this primitive cell (supercell) is larger than the overlayer unit cell and
contains several
atoms, then we will call this a high-order commensurate structure. When the
number of atoms in this cell goes to infinity, the high-order commensurate
structure becomes incommensurate.

With some exceptions, structures of strongly chemisorbed systems
are either fully commensurate (every adatom is in registry with the substrate
lattice) or incommensurate phases; while
high-order commensurate phases are less often seen among systems in which
strong adatom
localization occurs. Because of the localized nature of the adsorption and the
poor understanding of the interaction potentials, theoretical methods
to describe the structure and phase boundaries of these layers have relied by
and large on lattice gas models and the parametrization of interaction energies
\cite{einstein,binder}.

In the opposite limit, when incommensurate structures are formed, other
approaches have been tried. Calculations by Novaco and McTague
\cite{mctague}
and Shiba
\cite{shiba}
showed that orientational epitaxy might arise in an incommensurate
layer adsorbed on a substrate; this effect has been observed for rare-gas
monolayers adsorbed on graphite and for certain alkali monolayers on metal
substrates.

In this work we introduce a model that describes an ordered layer
with square symmetry adsorbed on a square substrate. This
model is used to compute the zero temperature phase diagram for this
system.

\section{Experimental Background}

Recently, high-order commensurate structures have been detected for Hg and Pb
adsorbed on Cu(001)
\cite{li2,li1}.
In addition to the fully commensurate $C(2 \times 2)$
phase, two high order commensurate square phases have been observed
and studied in details: the $C(4 \times 4)$ phase (Hg on Cu)
and the $(5 \times 5)R \tan^{-1} (3/4)$
phase (Pb on Cu). Only one of the atoms of the supercell is in the ideal
registry position, while all the
others sit in lesser energetically favorable locations. The experimental
results indicate that in these phases the adlayer forms an
almost perfect square lattice with only small local relaxations
caused by the substrate potential.

Surprisingly, these high-order commensurate phases possess square symmetry.
We recall that for an unsupported layer with
Lennard-Jones pairwise interactions the most stable structure has hexagonal
symmetry
\cite{biham}.
It is then clear that these high-order commensurate square structures must be
stabilized by substrate forces and/or by more complicated
interactions, such as three-body forces or angular dependent, covalent-like
interactions
\cite{einstein}.

\section{Classification of Square Phases}

We have developed a system, based on the two dimensional analog of the
Farey tree of rational numbers \cite{niven},
to classify
all the commensurate square phases on a square substrate.
Consider a square monolayer on a square substrate
such as the $(5 \times 5)R \tan^{-1} (3/4)$ phase (Fig. 1).
The unit cell of the monolayer is a square, which is typically
rotated with respect to the substrate main symmetry directions.
To characterize the structure we pick a side of the square and
project it on the x and y directions. Both projections are
integer multiples of the substrate lattice constant, which will
be given by the integers $p_x$ and $p_y$ respectively.
The linear size of the supercell,
in terms of the monolayer lattice constant,
is given by $q$.
Therefore, one can identify
each square phase by the pair of rational numbers
\begin{equation}
\left( {p_x \over q}, {p_y \over q} \right).
\end{equation}
Using this observation we
propose a systematic hierarchical procedure to generate a class of
commensurate square phases. The advantage of this method is that it
generates a sequence of phases from the most stable and lowest order ones up
to the least energetically favorable.

The list of phases is constructed by starting with a set of four
lowest order phases (furthest corners in Table 1).
Then between each pair of neighboring phases
one adds a new phase for which the new $p_x, p_y, q$ are the sums of the
corresponding quantities in the neighboring phases. The
four new phases are located at each side of the square
(midway between corners in Table 1).
A new phase is also introduced in the center, using as
input the two phases on one of the diagonals. Here one should choose
the diagonal that includes a phase which belongs to a lower generation.

One can continue this procedure by breaking the new
diagram into four separate smaller squares and repeating the procedure for
each one of them. The main advantage of this procedure is that
it generates phases of low order first (which are most likely to be
stable) while higher order phases appear in later generations.
This construction  should provide a
useful guide for the experimentalist willing to explore
interesting yet undetected structures which are in
the vicinity of experimentally observed ones.

\section{The  Model}

In order to obtain a better understanding of which adlayer structures one
should expect to find on a square substrate, we have studied a zero temperature
model and calculated its phase diagram.
In particular we consider a two dimensional layer of atoms
with Lennard-Jones interactions,
subject to a periodic underlying potential due to  interactions
with the substrate.
Since the Lennard-Jones interaction is of long range,
each atom in the overlayer interacts
with all the other atoms of the overlayer and the substrate.
The lattice constant, orientation and position with respect to the substrate
can all be changed to minimize the energy.
Since the square phases detected experimentally seem to show little relaxation
(and this is born out by  a mean-field type of calculation
\cite {biham}), one can consider that the adlayer
atoms are constrained to a perfect square structure.

In this simplified case one can use results for lattice
sums to add the adlayer-substrate interactions into an effective potential.
The energy will then be:
\begin{equation}
E=\sum_{m,n}^{\ \ \  \prime}
 V_{LJ} ( \vec r_{m} - \vec r_{n} ) + \sum_n V_S(\vec r _n)
\label{oned}
\end{equation}
where
\begin{equation}
V_{LJ} ( \vec r ) =
4 \varepsilon_{ad} \left[
  { \left( \sigma_{ad} \over \vert \vec r \vert \right)^{12}}
- { \left(  \sigma_{ad} \over \vert \vec r \vert \right)^{6}}
\right]
\label{lenjones}
\end{equation}
is the Lennard-Jones potential between the adlayer atoms,
$V_S$ is the underlying substrate potential
and $\vec r_n = (x_n,y_n)$
is the position of the nth adlayer atom.
The substrate potential is given by:
\begin{equation}
V_S(\vec r_n)=A_0 + \sum_{\vec k} A_{\vec k} e^{\vec k \cdot \vec r_n}
\label{fseries}
\end{equation}
where the sum runs over all the reciprocal lattice vectors of the substrate.
The Fourier coefficients are given by
\cite{steele}
\begin{equation}
A_0=8 \pi \varepsilon \sigma^2
\left[ {2 \over 5} \left( {\sigma \over z_0}\right)^{10} -
   \left( {\sigma \over z_0}\right)^4 \right]
\end{equation}
and
\begin{equation}
A_{\vec k}=
{ {8 \pi \varepsilon \sigma^{12}} \over { 30 \  a_{sb}^2 } }
\left( {k \over 2 z_0}\right)^{5} K_{5}(kz_0)
-{ { {8 \pi \varepsilon \sigma^6} \over { a_{sb}^2 } }
\left( {k \over 2 z_0}\right)^{2} K_{2}(kz_0)}
\label{fcoeff}
\end{equation}
where
$k=\vert \vec k \vert$,
$K_{p/q}$ is a modified    Bessel function of order p/q,
$z_0$ is the distance between the adlayer and substrate
and
$a_{sb}=(2 C_{12} / C_6)^{1/6} \cdot \sigma_{sb}$
is the lattice constant of the substrate
where
$C_n= {1 \over 2} \sum^{\prime}_{i,j} 1/(i^2+j^2)^{n/2}$.
Here, the parameters $\sigma$ and $\varepsilon$ for the adlayer-substrate
interaction are given by
$\sigma= ( \sigma_{ad} + \sigma_{sb} ) / 2$
and
$\varepsilon=\sqrt{ \varepsilon_{ad} \  \varepsilon_{sb} }$.
For simplicity we choose
$\varepsilon_{sb}=1$ and $\sigma_{sb}=1$.
The distance $z_0= \sigma$
between the adlayer and the substrate is then
determined so that the zeroth order term, $A_0$, in the potential
is minimized.
Note that we assume here that the distance $z_0$ between the
monolayer and the substrate is the same for all adlayer atoms, although
in reality these heights vary (0.1-0.2 $\AA$)
\cite{li0}
and depend on
the position of the atom with respect to the substrate.

To study the phase diagram we first choose a structure
and then calculate the energy per atom. We  compare these energies
for a set of commensurate and incommensurate square phases
and obtain the phase diagram for the
physically interesting regime
of $1<\sigma_{ad}<2$ (Fig. 2).
The  parameters in the phase diagram are $\sigma_{ad}$,
and $1/\sqrt{ \varepsilon_{ad} }$, the latter being
proportional to the strength of  the underlying substrate potential.
We  normalize the  lattice constant of the adlayer
in units of the substrate periodicity and define
$a=a_{ad}/a_{sb}$.
The phase diagram contains both commensurate
and incommensurate square phases.
Each commensurate phase is stable in a tongue
of a finite width,
in which
$a=\sqrt{p_x^2+p_y^2} \ /q$,
while incommensurate
phases exist in the gaps between these tongues.
In Fig. 2 one can see that the tongues start very narrow and
as the parameter
$1/\sqrt{ \varepsilon_{ad} }$
increases they become broader
until they intersect in triple points. Above the triple point
there is a first order transition line between the two tongues, while all
the incommensurate phases between them disappear.
The phase diagram is asymmetric and this is
due to the hard core nature of the Lennard-Jones potential.
The adlayer phase can expand considerably in order to gain
energy from the underlying substrate potential but it cannot contract as much
due to the hard core repulsion.
Also note that the high order tongues are extremely thin.
This is due to the asymptotic properties of the modified Bessel functions,
which appear in the Fourier coefficients $A_k$.
For high order harmonics of the potential the amplitudes
decay exponentially, or $A_k \propto e^{-z_0 k}$. Thus, the energy gain of high
order commensurate phases becomes rapidly very small as the order of
commensuration increases.

Finally, we have calculated the phase diagram with the inclusion of
(incommensurate) hexagonal phases; in
this case, most of the square phases are washed out, except for the fully
commensurate ones.  This is to be expected, since high order phases appear for
weak interactions with the substrate (lower part of Fig. 2) and
the energy gain for commensuration is small.

\section{Conclusion}
In summary, we have studied a simple model for the energetics of square
monolayer phases on a square substrate.
To study the phase diagram we introduced a simple classification
that allowed us to find the most energetically
favored phases. We find that when only square phases are considered,
each one of them has a tongue in which it is stable.
The $C(4 \times 4)$ phase (Hg on Cu)
and the
$(5 \times 5)R \tan^{-1} (3/4)$
phase (Pb on Cu) belong to very narrow tongues (not shown in Fig.2) which
intersect the $\sigma_{ad}$ axis at
$1.265$  and
$1.25$, respectively.
The fact that  hexagonal
phases win out (in our calculation) might be due to a variety of factors,
including the fact that the observed phases are metastable and the actual
interactions might be more complicated than the ones used in this study.

One of us (G.V.) would like to acknowledge partial support
from NSF grant 8802512.

\newpage

\baselineskip=24pt
\setlength{\parindent}{0em}
\setlength{\parskip}{1em}
\bigskip
\leftline{\underline{\bf \large Figure Captions}}

\noindent
Figure 1: The
$(5 \times 5)R \tan^{-1} (3/4)$
phase that appears in Pb on Cu(001). Empty circles represent
substrate atoms, while full circles represent adlayer atoms.

\noindent
Figure 2: Phase diagram (at 0 K) of two dimensional
          homogeneous square phases on a square
          substrate, with Lennard-Jones interactions.
          Note the clear asymmetric shape, which is due to the
          asymmetry of the Lennard-Jones potential.
It includes three fully commensurate square phases:
$1 \times 1$ (or $(0/1,1/1)$), $C(2 \times 2)$ (or $(1/1,1/1)$)
and $2 \times 2$ (or $(0/1,2/1)$). It also includes a few examples of the
infinitely many high order commensurate square phases (from left to
right):
$[1/2,2/2]$, $\{3/5,6/5\}$, $[0/2,3/2]$, $[1/2,3/2]$, and $\{3/4,7/4\}$.

\bigskip
\leftline{\underline{\bf \large Table Caption}}

\noindent
Table 1: A classification for square monolayer phases on a
square substrate. In this hierarchical classification, the
lowest order phases, which are typically the most stable, appear first,
while higher order phases appear in later generations. Generations are
distinguished by different symbols:
$( \  )$, $[ \ ]$, $\langle \  \rangle$, and $\{ \  \}$
for generations 1, 2, 3 and
4, respectively.
For example, the experimentally observed high order commensurate
square phases
$C(4 \times 4)$ and
$(5 \times 5)R \tan^{-1} (3/4)$
appear here as
$\{2/5,6/5\}$ and $\{3/4,4/4\}$ respectively.

\newpage
\topmargin=0.4in
\oddsidemargin=0.2in
\textwidth 38em
\textheight 105ex
\baselineskip=40pt
\pagestyle{empty}

\LARGE

\begin{table}
\begin{tabular}{rrrrrrrrr}
$\left( { 0 \over 1}, {1 \over 1} \right)$ &
                                           &
                                           &
                                           &
$\left[ { 0 \over 2}, {3 \over 2} \right]$ &
                                           &
                                           &
                                           &
$\left( { 0 \over 1}, {2 \over 1} \right)$ \\
                                           &
                                           &
                                           &
                                           &
                                           &
                                           &
                                           &
                                           &
                                           \\
                                           &
                                           &
                                           &
                                           &
                                           &
                                           &
                                           &
                                           &
                                           \\
                                           &
                                           &
                                           &
                                           &
                                           &
                                           &
                                           &
                                           &
                                           \\
$\langle { 1 \over 3}, {3 \over 3} \rangle$ &
                                           &
$\langle { 1 \over 3}, {4 \over 3} \rangle$ &
                                           &
                                           &
                                           &
                                           &
                                           &
                                           \\
                                           &
                                           &
                                           &
                                           &
                                           &
                                           &
                                           &
                                           &
                                           \\
                                           &
$\left\{ { 2 \over 5 }, { 6 \over 5 } \right\}$&
                                           &
                                           &
                                           &
                                           &
                                           &
                                           &
                                           \\
                                           &
                                           &
                                           &
                                           &
                                           &
                                           &
                                           &
                                           &
                                           \\
$\left[ { 1 \over 2}, {2 \over 2} \right]$ &
                                           &
$\langle { 2 \over 4}, {5 \over 4} \rangle$ &
                                           &
$\left[ { 1 \over 2}, {3 \over 2} \right]$ &
                                           &
                                           &
                                           &
$\left[ { 1 \over 2}, {4 \over 2} \right]$ \\
                                           &
                                           &
                                           &
                                           &
                                           &
                                           &
                                           &
                                           &
                                           \\
                                           &
                                           &
                                           &
                                           &
                                           &
                                           &
                                           &
                                           &
                                           \\
                                           &
                                           &
                                           &
                                           &
                                           &
                                           &
                                           &
                                           &
                                           \\
$\langle { 2 \over 3}, {3 \over 3} \rangle$ &
$\left\{ { 4 \over 6}, {7 \over 6} \right\}$ &
$\langle { 2 \over 3}, {4 \over 3} \rangle$ &
                                           &
                                           &
                                           &
                                           &
                                           &
                                           \\
                                           &
                                           &
                                           &
                                           &
                                           &
                                           &
                                           &
                                           &
                                           \\
$\left\{ { 3 \over 4}, {4 \over 4} \right\}$ &
$\left\{ { 3 \over 4}, {5 \over 4} \right\}$ &
$\left\{ { 5 \over 6}, {8 \over 6} \right\}$ &
                                           &
                                           &
                                           &
                                           &
                                           &
                                           \\
                                           &
                                           &
                                           &
                                           &
                                           &
                                           &
                                           &
                                           &
                                           \\
$\left( { 1 \over 1}, {1 \over 1} \right)$ &
$\left\{ { 4 \over 4}, {5 \over 4} \right\}$ &
$\langle { 3 \over 3}, {4 \over 3} \rangle$ &
\ \ \ \ \ \ \ \                            &
$\left[ { 2 \over 2}, {3 \over 2} \right]$ &
\ \ \ \ \ \ \                              &
\ \ \ \ \ \ \                              &
\ \ \ \ \ \ \                              &
$\left( { 1 \over 1}, {2 \over 1} \right)$ \\
\end{tabular}
\end{table}

\end{document}